\documentclass[showpacs,prb,twocolumn,floats,superscriptaddress]{revtex4}

\usepackage{graphicx}
\graphicspath{{Figures/}}
\usepackage{amsmath}
\usepackage{amssymb}
\usepackage[colorlinks=true,citecolor=blue,linkcolor=blue]{hyperref}
\usepackage{amsfonts}
\usepackage{url}
\usepackage{natbib}
\usepackage{color}
\usepackage{textcomp}

\renewcommand{\vec}[1]{\ensuremath{\boldsymbol{#1}}}

\begin{document}

\title{Klein tunneling through double barrier
in ABC-trilayer graphene}
\date{\today}
\author{Abderrahim El Mouhafid}
\email{elmouhafid.a@ucd.ac.ma}
\affiliation{Laboratory of Theoretical Physics, Faculty of Sciences, Choua\"ib Doukkali University, PO Box 20, 24000 El Jadida, Morocco}
\author{Ahmed Jellal}
\email{a.jellal@ucd.ac.ma}
\affiliation{Laboratory of Theoretical Physics, Faculty of Sciences, Choua\"ib Doukkali University, PO Box 20, 24000 El Jadida, Morocco}
\affiliation{Canadian Quantum Research Center, 204-3002 32 Ave Vernon, \\ BC V1T 2L7, Canada}
\author{Miloud Mekkaoui}
\email{miloud.mekkaoui@gmail.com}
\affiliation{Laboratory of Theoretical Physics, Faculty of Sciences, Choua\"ib Doukkali University, PO Box 20, 24000 El Jadida, Morocco}

\pacs{ 72.80.Vp, 73.21.Ac, 73.23.Ad}

\begin{abstract}
Klein tunneling and conductance for Dirac fermions in ABC-stacked trilayer graphene (ABC-TLG) through symmetric and asymmetric double potential barrier are investigated using the two and six-band continuum model. Numerical results for our system show that the transport is sensitive to the height, the width and the distance between the two barriers. Klein paradox at normal incidence and resonant features at non-normal incidence in the transmission result from resonant electron states in the wells or hole states in the barriers and strongly influence the ballistic conductance of the structures.
\end{abstract}

\maketitle
%
\section{Introduction}
Generally, graphene\cite{Novoselov2004,Novoselov2005,Zhang2005,Geim} is  a two dimensional (2D) lattice of carbon atoms arranged in hexagonal geometry. Its stacking  can be realized in different methods to engineer multi-layered graphene  showing  various physical properties. Typical examples of stacking includes Order, Bernal (AB), and Rhombohedral stacking (ABC)\cite{Aoki2007,Dresselhaus200286,Jhang11, Koshino2010,Lui11,Koshino200923}. In
the first one all carbon atoms of each layer are well-aligned, while  the second and third ones 
having  cycle periods composed of  two layers and three layers of non-aligned graphene, respectively. It has been showed  that 
the band structure, Klein tunneling, band gap, transport  and optical properties
of graphene  depend on the way how it is stacked \cite{Latil2006,Koshino2010,Mikito2013,Aoki2007,Cocemasov2013,Mak2010,Guinea200626,Avetisyan200901, Avetisyan201032,Lu200627,Ben201301,Chegel201683,Sheng200966,Celal2016} and 
the applied  external sources \cite{Aoki2007,Koshino2010,Kumar201101,Katsnelson2006,Peeters2010,Kumar2012,CommentBen,vanduppen205427,Ben201301, vanduppen195439,Pereira,Mirzakhani2017,ElMouhafid2017, Benlakhouy114835}.

One of the latest focuses a few of the multi-layer structures is the trilayer graphene (TLG)\cite{Guinea200626,Mirzakhani2017,Craciun2009,Kumar2012, vanduppen195439,Ben201301,Koshino2009,Zhang2010,Sena2011,Zhang2013,
Jung2013,SalahUddin2014,Marcos2014,Ma2012,Yuan2011,Menezes2014,
Yin2017,Barlas2012, Cote2012,
Xu2015,Yun2016,Kumar2014}, which  has two distinct allotropes such that 
 the Bernal (ABA) and Rhombohedral (ABC) stackings. 
 For ABA we have atoms of the top layer lie exactly on top of the bottom layer in addition to a dispersion relation as 
 a combination of the single layer linear energy band and the quadratic dispersion of bilayer graphene with no  opening gap under applied external electric field\cite{Lui2011}.
 %
 %
 While ABC has atoms of one of the sublattices of topmost layer lie above the center of the hexagons of  bottom layer 
 and shows  a  dispersion relation 
 approximately cubic with conduction and valence bands touching each other at a point close to the highly symmetric $\mathbf{K}$ and $\mathbf{K'}$ points\cite{Aoki2007,Zhang2010}. Consequently, the way in which the layers are stacked modifies strongly the energy spectrum of the resulted system and this of course affects its transport properties as well. Moreover, for TLG it was showed that  the emergence  of a Klein tunneling strongly depends on the staking order, being present only in ABC-TLG\cite{Kumar2012,CommentBen,Ben201301}. 
 Experimentally,
  the transport properties of ABC-
 and ABA-TLG 
 using dual locally gated field effect devices was
 investigated \cite{Zhang2013} where 
 an opening of a band gap was observed.
 The  giant conductance oscillations in ballistic TLG Fabry-Pérot interferometers was realized \cite{Campos2012},
 which came as a result from the phase coherent transport through resonant bound states subjected to an electrostatic barrier.


Motivated by different 
achievements listed above, especially  \cite{Kumar2012,CommentBen,Ben201301},
we study the Kelin tunneling of Dirac fermions in ABC-TLG scattered by 
a square double barrier. More precisely, the transmission probabilities
and conductance of electrons  will be investigated  by tacking into
account the full six band energy spectrum. We analyze two
interesting cases by making comparison between the incident
energy $E$ and interlayer coupling parameter $\gamma_1$. Indeed,
for $E<\gamma_1$ there is only one channel of transmission
exhibiting resonances, even for incident particles of energy $E$ less
than the 
barrier heights $V_j$, i.e.  $E< V_j$, depending on the
double barrier profile. For $E > \gamma_{1}$, we end up with tree
propagating modes resulted from nine possible ways of
transmission. Subsequently, we use the transfer matrix  and
density of current to determine all  transmission channels 
together with  the  conductance associated to our system. 
Under the appropriate choices of 
the physical
parameters characterizing our system, we numerically analyze 
our  results
and  compare them with literature.

The outline of the  present paper is as follows. In section \ref{Theory}, we
establish a mathematical framework using the full band model
to determine the eigenvalues and eigenvectors through double barrier. In section \ref{tunneling},
by matching the eigenspinors of different regions at interfaces and 
using the transfer matrix together with
the current density, we obtain nine channels of  transmission and reflections
as well as the associated 
conductance. We study 
 two band tunneling 
for symmetric and asymmetric  barrier  with and without the interlayer potential
difference. At high energy, we repeat  the same task  but by
considering full band  and emphasis what makes  difference with
respect to other case. In section \ref{Conductance}, we numerically study 
the basics features of 
 the resulted conductance and show the effect of symmetric and asymmetric barrier. Finally, we briefly summarize our main
findings. 

\section{Theory formula}
\label{Theory}
\subsection{Eigenvalues and eigenvectors}
\begin{figure}[tbh]
\begin{center}
\hspace{-0.25cm}\includegraphics[width=3.5in]{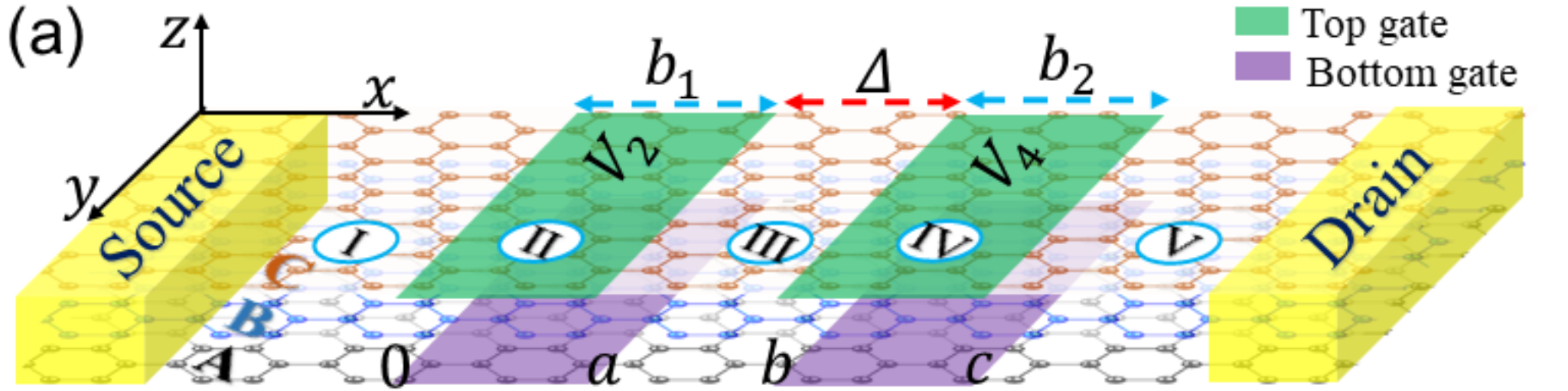}
\includegraphics[width=7.5cm]{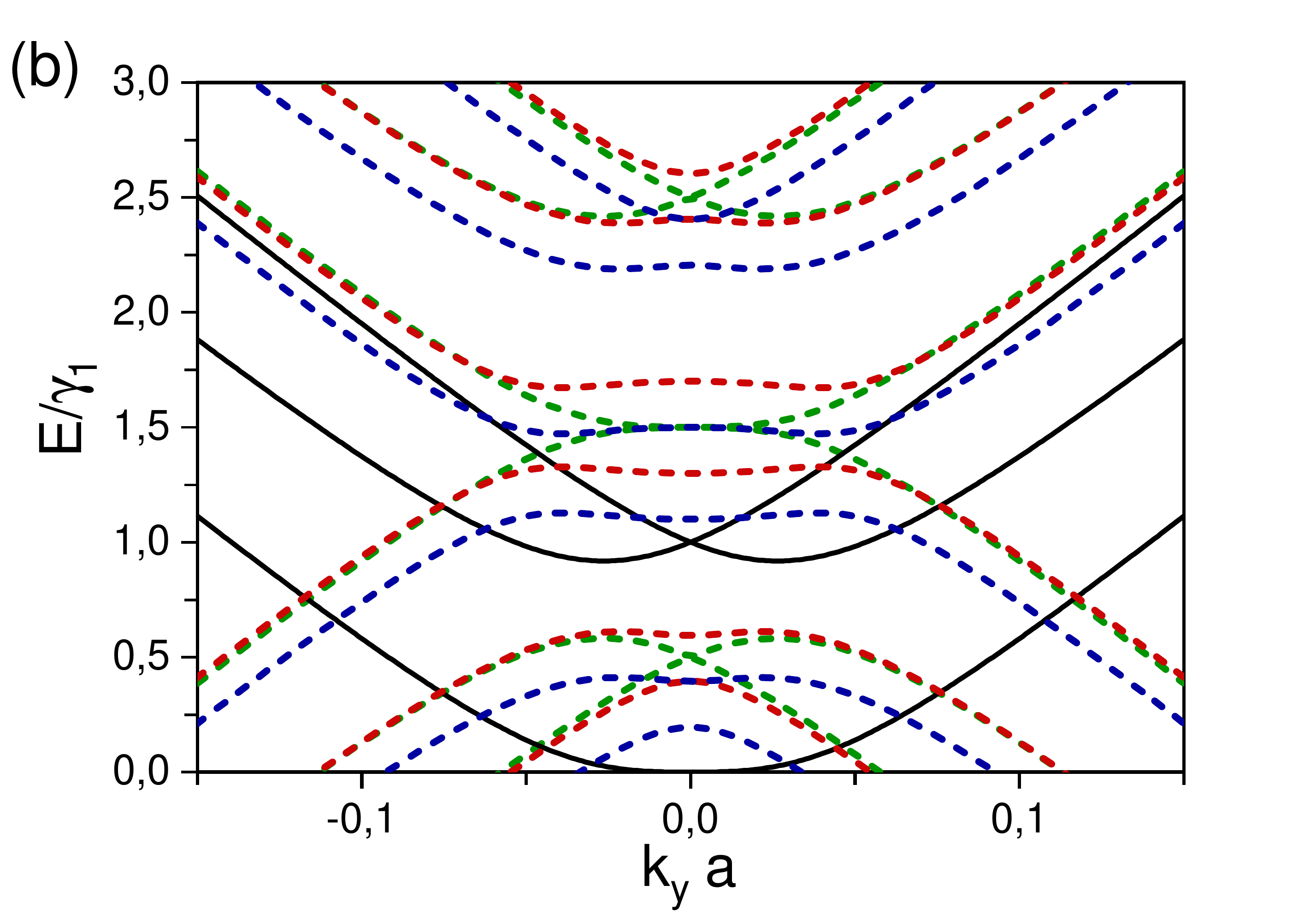}
\includegraphics[width=3.3in]{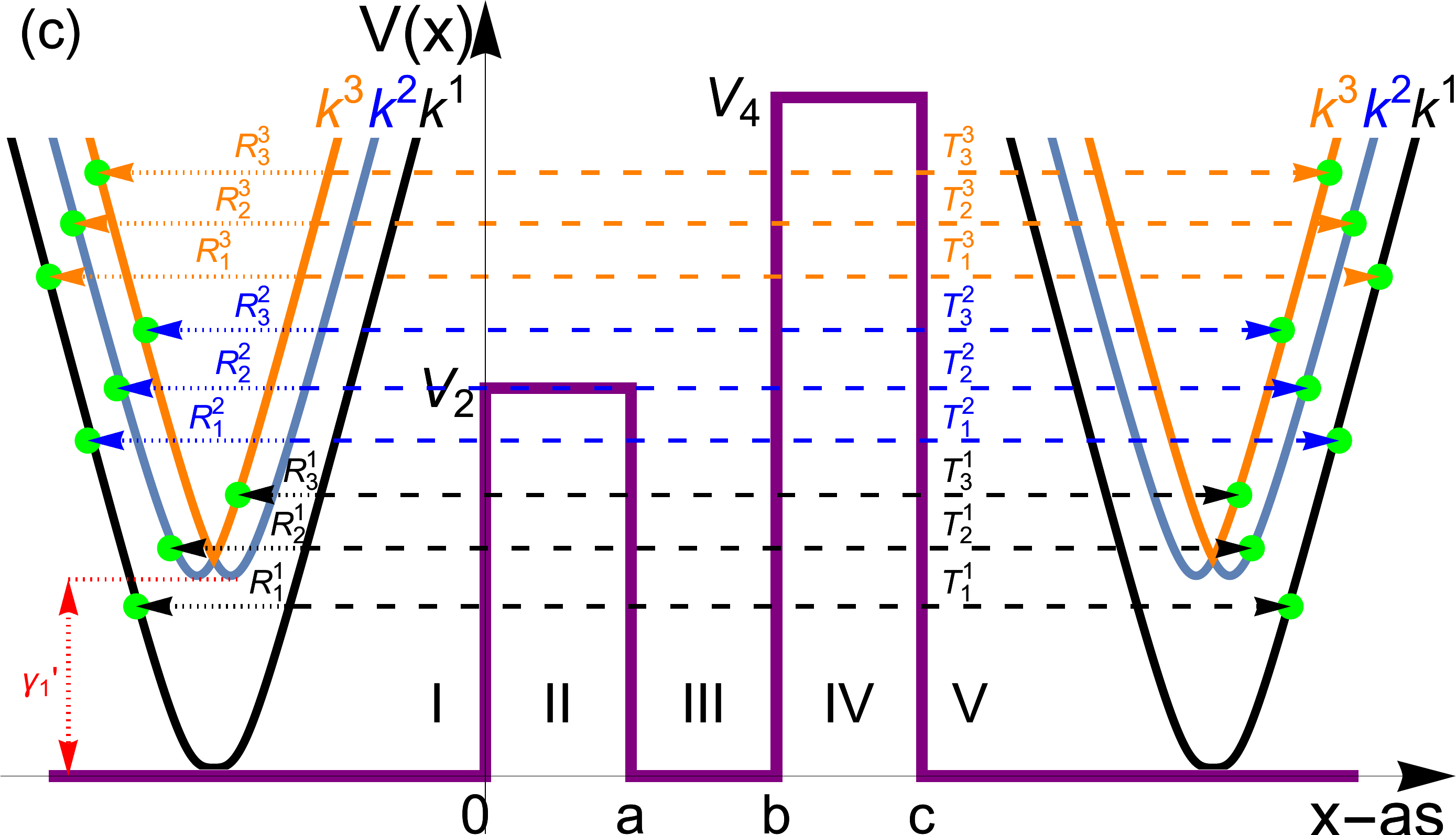}
\end{center}
 \caption{(Color online) Tunneling through a double barrier potential in ABC-TLG graphene. (a): Profile of a rectangular double barrier applied to our system.
 Indeed, an electrostatic potential difference $2\delta$ between top and bottom layers is applied in region II and IV. While, the electrostatic potential equal $V_2$ and $V_4$ for all three layers are applied in region II and IV, respectively. A, B, and C are three distinct positions of the hexagonal lattice when graphene trilayers are stacked. (b): Energy as function of the momentum $k_y$.  $V_2=V_4=1.5\gamma_1$ and $\delta_2=\delta_4=0$ (green lines).  $V_2=V_4=1.5\gamma_1$ and $\delta_2=\delta_4=0.2\gamma_{1}$ (red lines).  $V_2=1.3\gamma_1$, $V_4=1.5\gamma_1$ and $\delta_2=\delta_4=0.2\gamma_1$ (blue lines). The dashed and thick lines represent the band inside and outside the barriers, respectively. (c): Schematic representation of  different modes as well as the corresponding transmission and reflection channels. The dots and arrows indicate the energy regions for which electrons impinging perpendicularly on the barrier will be transmitted or reflected. We have chosen $\gamma'_1= 0.918 \gamma_{1}.$ }\label{figbarriermod}
\end{figure}

Single layer graphene (SLG) has a hexagonal crystal structure of two sublattices 
A and B with 
the interatomic distance  $a=0.142$ nm \cite{Partoens2007} and the intra-layer coupling  $\gamma_0\approx3$ eV\cite{Zhang2011}. On the other hand, TLG graphene   is a tree stacked SLG (Rhombohedral stacking) and has an unit cell 
of  six atoms\cite{Cote2013}. 
Under the nearest-neighbor tight binding approximation, one can derive the 
 Hamiltonian describing ABC-TLG near the $\mathbf{K}$ point \cite{Zhang2010}
\begin{equation}\label{eqHam}
H=\hbar v_{F}
\begin{pmatrix}
\vec{\sigma}\cdot\vec{k} & \Gamma & 0 \\
\Gamma ^{\dag } & \vec{\sigma}\cdot\vec{k} & \Gamma \\
0 & \Gamma ^{\dag } & \vec{\sigma}\cdot\vec{k}%
\end{pmatrix} ,
\end{equation}%
 in the basis of the atomic orbital eigenfunctions
\begin{equation}
	\Psi =\left( \psi _{A_{1}},\psi _{B_{1}},\psi _{A_{2}},\psi _{B_{2}},\psi_{A_{3}},\psi _{B_{3}}\right)^{T} ,  \label{}
\end{equation}
with $\vec{\sigma}=(\sigma_x,\sigma_y)$ a vector of Pauli matrices, the Fermi velocity $v_{F}=3a_0\gamma_0/(2\hbar)$,  
 the wave vector $\vec{k}$ and $\Gamma$ 
 is the interlayer coupling
\begin{equation}
\Gamma =\frac{1}{\hbar v_{F}}
\begin{pmatrix}
0 & 0 \\
\gamma _{1} & 0
\end{pmatrix},
\end{equation}
where $\gamma_{1}=0.4$ eV is the nearest neighbor coupling term between adjacent layers. 
Based on  
the profile of double barrier, see Fig. \ref{figbarriermod}(a),
we set all regions composing our system as
$j=\text{I}$
 ($x\leq 0$), $j=\text{II}$  ($0<x\leq a$), $j=\text{III}$   ($a<x\leq b$), $j=\text{IV}$
 ($b<x\leq c$) and $j=\text{V}$   ($x>c$) where $a=b_1$, $b=b_1+\Delta$ and $c=b_1+\Delta+b_2$. Thus,  in the $j$-th region (\ref{eqHam}) takes the form
\begin{equation}\label{Hamj}
\mathcal{H}_j=
\begin{pmatrix}
  V_j+\delta_j & \hslash v_{F}\pi^{\dag} & 0 & 0 & 0 & 0 \\
  \hslash v_{F}\pi &  V_j+\delta_j & \gamma_{1} & 0 & 0 & 0\\
  0 & \gamma_{1} & V_j & \hslash v_{F}\pi^{\dag} & 0 & 0 \\
  0 & 0& \hslash v_{F}\pi & V_j & \gamma_{1} & 0 \\
    0 & 0& 0 & \gamma_{1} &  V_j-\delta_j & \hslash v_{F}\pi^{\dag} \\
      0 & 0& 0 & 0 & \hslash v_{F}\pi  &  V_j-\delta_j \\
\end{pmatrix}
\end{equation}
where $\pi=p_{x}+ip_{y},
\pi^{\dag}=p_{x}-ip_{y}$ are the in-plan momenta and its conjugate
with $p_{x,y}=-i\hbar\partial_{x,y}$. $V_j$ is the electrostatic potential equal for all three layers applied in region II and IV and $\delta_j$ is the interlayer potential difference between top and bottom layers applied in region II and IV. In regions I, III and V we have $V_j=\delta_j=0$. 
Since the Hamiltonian \eqref{Hamj} commutes with the momentum $p_y$, then we can proceed by separating the eigenspinors as
\begin{equation}
\psi^j(x,y) =e^{ik_y y}[{\phi}^j_{A_{1}
},{\phi}^j_{B_{1}},{\phi}^j_{A_{2}},{\phi}^j_{B_{2}},{\phi}^j_{A_{3}},{\phi}^j_{B_{3}}]^{\dag}, \label{wavef}
\end{equation}
where $\dag$ stands for the transpose of the row vector.
According to Fig. \ref{figbarriermod}(a), we  have basically two different sectors
with zero (I, III, V) and nonzero (II, IV) potentials. Then, we can derive  a general
solution  in the second sector and require  
$V_{j}=\delta_j=0$ to find that for the 
first one. To simplify our notation, we  introduce the length
scale $l=\frac{\hbar v_{F}}{\gamma_{1}}\approx 1.64 \ nm$, which represents the interlayer coupling length,
and define $E_j \rightarrow\ \frac{E}{\gamma_1}$, $V_j \rightarrow\ \frac{V_j}{\gamma_1}$, $\delta_j \rightarrow\ \frac{\delta_j}{\gamma_1}$, $k_y \rightarrow\ lk_y$, $\vec r \rightarrow\ \frac{\vec r}{l}$.

As usual, to derive the eigenvalues and the eingespinors we solve
$H_j\psi_j=E_j\psi_j$. Then, by replacing Eqs. (\ref{Hamj}) and (\ref{wavef}) we obtain six coupled differential equations
\begin{subequations}\label{sys1}
 \begin{align}
&-i(\partial_{x}+k_{y})\phi^j_{B_{1}} =(\varepsilon_j-\delta_j)\phi^j_{A_{1}},
 \label{eqsd1}\\
&-i(\partial_{x}-k_{y})\phi^j_{A_{1}} =(\varepsilon_j-\delta_j)\phi^j_{B_{1}}-\phi^j_{A_{2}}, \label{eqsd2}  \\
&-i(\partial_{x}+k_{y})\phi^j_{B_{2}}=\varepsilon_j\phi^j_{A_{2}}-\phi^j_{B_{1}}, \label{eqsd3}  \\
&-i(\partial_{x}-k_{y})\phi^j_{A_{2}}=\varepsilon_j\phi^j_{B_{2}}-\phi^j_{A_{3}},
\label{eqsd4} \\
&-i(\partial_{x}+k_{y})\phi^j_{B_{3}}=(\varepsilon_j+\delta_j)\phi^j_{A_{3}}-\phi^j_{B_{2}},
\label{eqsd5} \\
&-i(\partial_{x}-k_{y})\phi^j_{A_{3}}=(\varepsilon_j+\delta_j)\phi^j_{B_{3}},
 \label{eqsd6}
\end{align}
\end{subequations}
with $ \epsilon_j=E_j-V_{j}$ and 
 the wave vector $k_{y}$ along the $y$-direction. 
 We proceed further by decoupling the above set of equations.  For instance, we express 
 Eqs. \eqref{eqsd1} and \eqref{eqsd6}
 as
\begin{subequations}\label{}
\begin{align}
&\phi^j_{A_1}=-\frac{i}{\varepsilon_j-\delta_j}(\partial_{x}+k_{y})\phi^j_{B_{1}},\\ &\phi^j_{B_3}=-\frac{i}{\varepsilon_j+\delta_j}(\partial_{x}-k_{y})\phi^j_{A_{3}}.
\end{align}
\end{subequations}
which can be injected into Eqs. \eqref{eqsd2} and \eqref{eqsd5} to end up with two equations
\begin{subequations}\label{}
 \begin{align}
&(\partial^2_{x}-k^2_{y})\phi^j_{B_{1}}+(\varepsilon_j-\delta_j)\phi^j_{A_{2}}=(\varepsilon_j-\delta_j)^2\phi^j_{B_{1}},
\label{eq9a} \\
&(\partial^2_{x}-k^2_{y})\phi^j_{A_{3}}-(\varepsilon_j+\delta_j)\phi^j_{B_{2}}=-(\varepsilon_j+\delta_j)^2\phi^j_{A_{3}}.
\label{eq9b}
\end{align}
\end{subequations}
Now  substituting
Eqs. \eqref{eq9a} and \eqref{eq9b}, respectively, 
in Eqs. \eqref{eqsd3} and \eqref{eqsd4},
then after
a straightforward  algebra and
by setting 
\begin{subequations}\label{}
	\begin{align}
		&a_j=3\varepsilon_j^2+2\delta_j^2,
		\label{} \\
		&b_j=(\delta_j+\varepsilon_j)(3\varepsilon_j^3+3\delta_j^3+(\delta_j^2-2)\varepsilon_j+\delta_j\varepsilon_j^2),
		\label{}\\
		&c_j=(\delta_j^{^{2}}-\varepsilon_j^2)(1+\delta_j^2-\varepsilon_j^2)(1+\delta_j+\varepsilon_j)(-1+\delta_j+\varepsilon_j).
		\label{}
	\end{align}
\end{subequations}
one finds a sixth-order differential equation for $\phi^j_{A_{3}}$
\begin{widetext}
\begin{align}\label{eqcub}
\left[\frac{d^{6}}{dx^{6}}+\left(a_j-3k_y^2\right)\frac{d^{4}}{dx^{4}}-\left(2a_jk_y^2-3k_y^4-b_j\right)\frac{d^{2}}{dx^{2}}-k_y^6+a_jk_y^4-b_jk_y^2+c_j\right]\phi^j_{A_{3}}=0,
\end{align}
\end{widetext}

We show that 
the solution of Eq. \eqref{eqcub} can be expressed as a linear combination of plane waves, such as
\begin{equation}
\phi^j_{A_{3}}=\sum^{3}_{n=1}\left(a_{n}e^{ik^j_n x}+b_{n}e^{-ik^j_n x}\right),
\end{equation}
where $a_{n}$ and $b_n$ $(n = 1, 2, 3)$ are coefficients of normalization. The wave vectors $k^j$ along the $x$-direction in each region  are  solutions of the cubic equation
\begin{equation}\label{cubeq}
\left[(k^j)^{2}+k^2_y\right]^{3}-h^j_1\left[(k^j)^{2}+k^2_y\right]^{2}+h^j_2\left[(k^j)^{2}+k^2_y\right]-h^j_3=0,
\end{equation}
with the involved parameters
\begin{subequations}\label{}
\begin{align}
&h^j_1=(\alpha^j_1)^{2}+(\alpha^j_2)^{2}+(\alpha^j_3)^{2},
\label{} \\
& h^j_2=(\alpha^j_1\alpha^j_2)^{2}+(\alpha^j_1\alpha^j_3)^{2}+(\alpha^j_2\alpha^j_3)^{2}-\alpha^j_2(\alpha^j_1+\alpha^j_3),
\label{}\\
&
h^j_3=(\alpha^j_1\alpha^j_2\alpha^j_3)^{2}-\alpha^j_1\alpha^j_2\alpha^j_3(\alpha^j_1+\alpha^j_3)+\alpha^j_1\alpha^j_3,
\label{}
\end{align}
\end{subequations}
and $\alpha^j_1=\varepsilon_j+\delta_j$, $\alpha^j_2=\varepsilon_j$, $\alpha^j_3=\varepsilon_j-\delta_j$.

The rest of spinor components is given by
\begin{subequations}\label{}
 \begin{align}
&\phi^j_{B_{3}}=\sum^{3}_{n=1}\frac{1}{\alpha^j_1}\left(g^j_na_{n}e^{ik^j_n x}-f^j_nb_{n}e^{-ik^j_n x}\right),
\label{} \\
&\phi^j_{B_{2}}=\sum^{3}_{n=1}\frac{\rho^j_n}{\alpha^j_1}\left(a_{n}e^{ik^j_n x}+b_{n}e^{-ik^j_n x}\right),
\label{} \\
&\phi^j_{A_{2}}=\sum^{3}_{n=1}\frac{\mu^j_n}{\alpha^j_1}\left(\frac{a_n}{g^j_n}e^{ik^j_n x}-\frac{b_{n}}{f^j_n}e^{-ik^j_n x}\right),
\label{}\\
&\phi^j_{B_{1}}=\sum^{3}_{n=1}\frac{\lambda^j_n}{\alpha^j_1}\left(\frac{a_n}{g^j_n}e^{ik^j_n x}-\frac{b_{n}}{f^j_n}e^{-ik^j_n x}\right),
\label{}\\
&\phi^j_{A_{1}}=\sum^{3}_{n=1}\frac{\lambda^j_n}{\alpha^j_1\alpha^j_3}\left(\frac{f^j_n}{g^j_n}a_ne^{ik^j_n x}+\frac{g^j_n}{f^j_n}b_{n}e^{-ik^j_n x}\right),
\label{}
\end{align}
\end{subequations}
where $g^j_n=k^j_n+ik_y$, $f^j_n=k^j_n-ik_y$, $\rho^j_n=(\alpha^j_1)^{2}-(k^j_n)^{2}-(k_y)^{2}$, $\eta^j_n=(\alpha^j_2)^{2}-(k^j_n)^{2}-(k_y)^{2}$, $\mu^j_n=\alpha^j_2\rho^j_n-\alpha^j_1$, $\lambda^j_n=\rho^j_n\eta^j_n-\alpha^j_1\alpha^j_2$.  

The energy spectrum of our Hamiltonian \ref{Hamj} in the different regions are depicted in Fig. \ref{figbarriermod}(b). The spectrum consists of six energy bands symmetric at $E=0$ of which two touch each other at $k=0$. The dashed (solid) curves correspond the energy spectrum of ABC-TLG inside (outside) the barriers. We notice that when the ABC-TLG is subject to the inter-layer bias $\delta\neq0$ the tree bands are switched and a band gap is opening between them at the smallest potential $V_2$ or $V_4$ for the case when we have $V_2\neq V_4$ or at the potential $V_2$ when we have $V_2=V_4$. Indeed for $E<\gamma_{1}$, we have just one mode of propagation correspond the wave vector $k_1$, however when  $E>\gamma_{1}$, we have tree mode of propagation correspond the wave vectors $k_1$, $k_2$ and $k_3$ which presenting a new propagation mode which will be depicted in detail in Fig. \ref{figbarriermod}(c).

Now, we can write the general solution in each region as
\begin{equation}\label{eq9}
\psi^j(x,y)=Q^j M^j(x)C^j e^{ik_{y}y}
\end{equation}
in terms of the matrices
\begin{equation}\label{MatrixG}
Q^j=
\begin{pmatrix}
   \frac{\lambda^j_1f^j_1}{\alpha^j_1\alpha^j_3g^j_1} & \frac{\lambda^j_1g^j_1}{\alpha^j_1\alpha^j_3f^j_1} & \frac{\lambda^j_2f^j_2}{\alpha^j_1\alpha^j_3g^j_2} & \frac{\lambda^j_2g^j_2}{\alpha^j_1\alpha^j_3f^j_2} & \frac{\lambda^j_3f^j_3}{\alpha^j_1\alpha^j_3g^j_3} & \frac{\lambda^j_3g^j_3}{\alpha^j_1\alpha^j_3f^j_3} \\
  \frac{\lambda^j_1}{\alpha^j_1g^j_1} & -\frac{\lambda^j_1}{\alpha^j_1f^j_1} & \frac{\lambda^j_2}{\alpha^j_1g^j_2} & -\frac{\lambda^j_2}{\alpha^j_1f^j_2} & \frac{\lambda^j_3}{\alpha^j_1g^j_3} & -\frac{\lambda^j_3}{\alpha^j_1f^j_3}\\
  \frac{\mu^j_1}{\alpha^j_1g^j_1} & -\frac{\mu^j_1}{\alpha^j_1f^j_1} & \frac{\mu^j_2}{\alpha^j_1g^j_2} & -\frac{\mu^j_2}{\alpha^j_1f^j_2} & \frac{\mu^j_3}{\alpha^j_1g^j_3} & -\frac{\mu^j_3}{\alpha^j_1f^j_3} \\
  \frac{\rho^j_1}{\alpha^j_1} & \frac{\rho^j_1}{\alpha^j_1} & \frac{\rho^j_2}{\alpha^j_1} & \frac{\rho^j_2}{\alpha^j_1} & \frac{\rho^j_3}{\alpha^j_1} & \frac{\rho^j_3}{\alpha^j_1} \\
  1 & 1 & 1 & 1 & 1 & 1 \\
  \frac{g^j_1}{\alpha^j_1} & -\frac{f^j_1}{\alpha^j_1} & \frac{g^j_2}{\alpha^j_1} & -\frac{f^j_2}{\alpha^j_1} & \frac{g^j_3}{\alpha^j_1} & -\frac{f^j_3}{\alpha^j_1} \\
\end{pmatrix}
\end{equation}
\begin{equation}\label{}
M^j=
\begin{pmatrix}
  e^{ik^j_1 x} & 0 & 0 & 0 & 0 & 0 \\
  0 & e^{-ik^j_1 x} & 0 & 0 & 0 & 0\\
  0 & 0 & e^{ik^j_2 x} & 0 & 0 & 0 \\
  0 & 0 & 0 & e^{-ik^j_2 x} & 0 & 0 \\
  0 & 0 & 0 & 0 & e^{ik^j_3 x} & 0 \\
  0 & 0 & 0 & 0 & 0 & e^{-ik^j_3 x} 
\end{pmatrix}
\end{equation}
\begin{equation}\label{Cjjeq}
C^j=
\begin{pmatrix}
  a^{+}_1 \\
  b^{-}_1 \\
  a^{+}_2 \\
  b^{-}_2 \\
  a^{+}_3 \\
  b^{-}_3 \\
\end{pmatrix},
\end{equation}
where the subscript $i$ in Eq. \ref{Cjjeq} refers to the corresponding wave vector and the superscript plus/minus indicates the right/left propagation or evanescent states.  Since we are using the transfer matrix approach, 
 we are interested in the
normalization coefficients  (components of $C$).
For this purpose, we specify
our spinors in region I
\begin{subequations}\label{}
\begin{align}
&\phi^{\text{I}}_{A1}=\sum^{3}_{j=1}\left(\delta_{s,j}Q^1_{1,2j-1}e^{ik^{1}_jx}+r^{s}_jQ^1_{1,2j}e^{-ik^{1}_jx}\right)\\
&\phi^{\text{I}}_{B1}=\sum^{3}_{j=1}\left(\delta_{s,j}Q^1_{2,2j-1}e^{ik^{1}_jx}+r^{s}_jQ^1_{2,2j}e^{-ik^{1}_jx}\right)\\
&\phi^{\text{I}}_{A2}=\sum^{3}_{j=1}\left(\delta_{s,j}Q^1_{3,2j-1}e^{ik^{1}_jx}+r^{s}_jQ^1_{3,2j}e^{-ik^{1}_jx}\right)\\
&\phi^{\text{I}}_{B2}=\sum^{3}_{j=1}\left(\delta_{s,j}Q^1_{4,2j-1}e^{ik^{1}_jx}+r^{s}_jQ^1_{4,2j}e^{-ik^{1}_jx}\right)\\
&\phi^{\text{I}}_{A3}=\sum^{3}_{j=1}\left(\delta_{s,j}Q^1_{5,2j-1}e^{ik^{1}_jx}+r^{s}_jQ^1_{5,2j}e^{-ik^{1}_jx}\right)\\
&\phi^{\text{I}}_{B3}=\sum^{3}_{j=1}\left(\delta_{s,j}Q^1_{6,2j-1}e^{ik^{1}_jx}+r^{s}_jQ^1_{6,2j}e^{-ik^{1}_jx}\right)
\end{align}
\end{subequations}
and region V
\begin{subequations}\label{}
\begin{align}
&\phi^{\text{V}}_{A1}=t^s_1Q^{\text{V}}_{1,1}e^{ik^{1}_1x}+t^{s}_2Q^{\text{V}}_{1,3}e^{ik^{1}_2x}+t^{s}_3Q^{\text{V}}_{1,5}e^{ik^{1}_3x}\\
&\phi^{\text{V}}_{B1}=t^s_1Q^{\text{V}}_{2,1}e^{ik^{1}_1x}+t^{s}_2Q^{\text{V}}_{2,3}e^{ik^{1}_2x}+t^{s}_3Q^{\text{V}}_{2,5}e^{ik^{1}_3x}\\
&\phi^{\text{V}}_{A2}=t^s_1Q^{\text{V}}_{3,1}e^{ik^{1}_1x}+t^{s}_2Q^{\text{V}}_{3,3}e^{ik^{1}_2x}+t^{s}_3Q^{\text{V}}_{3,5}e^{ik^{1}_3x}\\
&\phi^{\text{V}}_{B2}=t^s_1Q^{\text{V}}_{4,1}e^{ik^{1}_1x}+t^{s}_2Q^{\text{V}}_{4,3}e^{ik^{1}_2x}+t^{s}_3Q^{\text{V}}_{4,5}e^{ik^{1}_3x}\\
&\phi^{\text{V}}_{A3}=t^s_1Q^{\text{V}}_{5,1}e^{ik^{1}_1x}+t^{s}_2Q^{\text{V}}_{5,3}e^{ik^{1}_2x}+t^{s}_3Q^{\text{V}}_{5,5}e^{ik^{1}_3x}\\
&\phi^{\text{V}}_{B3}=t^s_1Q^{\text{V}}_{6,1}e^{ik^{1}_1x}+t^{s}_2Q^{\text{V}}_{6,3}e^{ik^{1}_2x}+t^{s}_3Q^{\text{V}}_{6,5}e^{ik^{1}_3x},
\end{align}
\end{subequations}
where $\delta_{s,i}$ $(i=1,2,3)$ is the Kronecker delta, $s=1, 2 ,3$ indicate the mode of propagation or evanescent waves characterized by three different wave vectors $k_1$, $k_2$ and $k_3$. $Q^j_{l,m}$ are the elements of the matrix \eqref{MatrixG}.
 
 In regions I, III and V there is potential, then we immediately derive
  the
relation 
\begin{equation}\label{eq135}
Q^{\text{I}} M^{\text{II}}(x)=Q^{\text{III}} M^{\text{III}}(x)=Q^{V}
M^{V}(x),
\end{equation}
analogue to Eq. \eqref{eq9}. These results will be used to deal with different issues related to our system. Indeed, we will compute all channels of transmissions
and reflections together with the associated conductance.
\subsection{Transmission  probabilities and conductance}
\begin{figure*}[tbh]
	\centering
	\includegraphics[width=7 in]{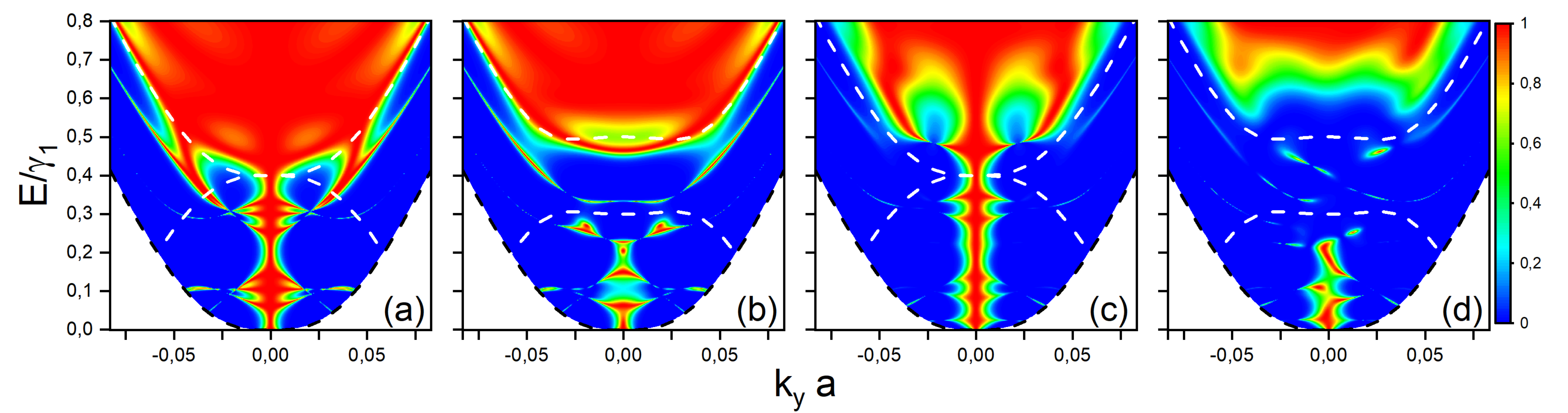}
	\caption{(Color online) Density plot of transmission probability $T^1_1$, for $\delta_2=\delta_4=0$ (a,c) and for $\delta_2=\delta_4=0.2\ \gamma_1$ (b,d), versus the incident energy $E$ and the wave vector $ky$ for $V_{2}=V_{4}=0.4\ \gamma_{1}$ (a,b) and for $V_{2}=0.4\ \gamma_{1}$, $V_{4}=0.6\ \gamma_{1}$ (c,d), with $b_1=b_2=\Delta=10$ nm. The dashed white lines show the band inside the barrier
		whereas 
		the black
		lines represent the band outside  the barrier.}\label{TransbasseE}
\end{figure*}
To determine the transmission and reflection we use the boundary conditions
in addition to the current density. We start from 
the continuity of the spinors at different interfaces  to obtain
the
coefficients in the incident and reflected regions
\begin{equation}\label{eq17}
C^{s}_{\text{I}}=\left(%
\begin{array}{cccccc}
 \delta_{s,1} \\
  r^{s}_{1} \\
  \delta_{s,2} \\
  r^{s}_{2} \\
   \delta_{s,3} \\
  r^{s}_{3} \\
\end{array}%
\right), \qquad
C^{s}_{\text{V}}=\left(%
\begin{array}{c}
  t^{s}_{1} \\
  0 \\
  t^{s}_{2} \\
  0 \\
  t^{s}_{3} \\
  0 \\
\end{array}%
\right)
\end{equation}
which can be coupled via
the transfer matrix $M$ as
\begin{equation}\label{eq18}
C^{s}_{\text{I}}=M C^{s}_{\text{V}}
\end{equation}
resulted from 
the continuity of spinors at the
four interfaces of the double barrier structure (Fig. \ref{figbarriermod}(a)). The remaining relations  are given by
\begin{align}\label{eq19}
    &Q^{\text{I}} M^{\text{I}}(0)C^{\text{I}}=Q^{\text{II}} M^{\text{II}}(0)C^{\text{II}},\\
    &Q^{\text{II}} M^{\text{II}}(a)C^{\text{II}}=Q^{\text{III}} 
    M^{\text{III}}(a)C^{\text{III}},\\
    &Q^{\text{III}} M^{\text{III}}(b)C^{\text{III}}=Q^{\text{IV}} M^{\text{IV}}(b)C^{\text{IV}},\\
    &Q^{\text{IV}} M^{\text{IV}}(c)C^{\text{IV}}=Q^{\text{V}} M^{\text{V}}(d)C^{\text{V}}.
\end{align}
Now solving the above system of equations and taking into account of Eq. \eqref{eq135}, one can find
the form of $M$.
Then we can specify the complex coefficients of the transmission
$t^{s}_{i}$ $(i=1, 2, 3)$ and reflection $r^{s}_{i}$ by using $M$.

To obtain the transmission $T$ and reflection $R$
probabilities, we introduce  the current density $\overrightarrow{\mathbf{J}}$
associated to our system.
Then, one can show
\begin{equation}\label{dcurm}
\overrightarrow{\mathbf{J}}=v_{F}\mathbf{\Psi}^{\dagger}\overrightarrow{\alpha}\mathbf\Psi,
\end{equation}
where $\vec{\alpha}$ is a $6\times6$  matrix having 
three Pauli matrices $\sigma_{x}$ in diagonal and the remaining elements are nulls.
It is clearly seen that Eq. \eqref{dcurm}
allows to obtain the incident $  \mathbf{J}_{\sf inc} $, 
transmitted $ \mathbf{J}_{\sf tra} $ and reflected $ \mathbf{J}_{\sf ref} $
density currents. Consequently, 
we get 
\begin{equation}\label{eq21}
T=\frac{|\mathbf{J}_{\sf tra}|}{|\mathbf{J}_{\sf inc}|}, \qquad
R=\frac{|\mathbf{J}_{\sf ref}|}{|\mathbf{J}_{\sf inc}|}.
\end{equation}
By using Eq. \eqref{eq9}, we explicitly determine $T$ and $R$
\begin{equation}\label{eq23}
T^{s}_{i}=\frac{A^x_{i,i}}{A^x_{s,s}}|t^{s}_{i}|^{2},\qquad
R^{s}_{i}=\frac{A^x_{i,i}}{A^x_{s,s}}|r^{s}_{i}|^{2},
\end{equation}
such that  $A^x_{i,i}$ and $A^x_{s,s}$ are  elements of the diagonal matrix $\overrightarrow{\mathbf{A}}=Q^{\dagger}\overrightarrow{\alpha}Q$ consisting of traceless $2\times2$ blocks where each one corresponds to a propagation mode $s$. These expressions can be explained as follows. Since we have six band, the electrons can be scattered between them and then we need to take into account the change in their velocities. With that, we find nine channels in transmission and reflection corresponding to  three modes of propagation  $k^1$, $k^2$ and $k^3$  solutions of the cubic Eq. \eqref{cubeq}. 

Fig. \ref{figbarriermod}(c) shows the modes $k^1$, $k^2$,  $k^3$ with theirs  transmission and reflection
probabilities through double barrier structure. 
For $E < \gamma_1$ (low energies ), there is only  the mode of propagation $k^1$, which  gives rise to one transmission $T^1_1$ and one reflection $R^1_1$ channel through the two conduction bands touching at zero energy on the both sides of the double barrier. For 
 $E > \gamma_1$ (higher energy), the three modes of propagation $k^1$, $k^2$, $k^3$ are present, which leads to  nine transmission $T^{1}_{i}$, $T^{2}_{i}$, $T^{3}_{i}$ $(i=1, 2, 3)$ and reflection $R^{1}_{i}$, $R^{2}_{i}$, $R^{3}_{i}$ channels, through the six conduction bands. For the transmission, there are three non scattered channels denoted by  $T^{1}_{1}$, $T^{2}_{2}$, $T^{3}_{3}$ for propagation via $k^1$, $k^2$, $k^3$, respectively, in addition to six scattered channels in which the particle enters via one channel and exits via another one.
 These will be specified as  $T^{1}_{3,2}$, $T^{2}_{3,1}$ and $T^{3}_{2,1}$ for scattering from the $k^1$ band to the $k^{3,2}$, from the $k^2$ band to the $k^{3,1}$ and from the $k^3$ band to the $k^{2,1}$ bands, respectively. Also we will adopt  the same definition for the  reflection channels, i.e. $R^{1,2,3}_{1,2,3}$. The eighteen channels are schematically depicted in Fig. \ref{figbarriermod}(c). We not that the transmission $T^{1,2,3}_{1,2,3}$ and reflection $R^{1,2,3}_{1,2,3}$ probabilities obeying the following equation 
\begin{equation}
\sum_{j=1,2,3}\left(T^j_{1,2,3}+R^j_{1,2,3}\right)=1.
\end{equation}
For example, for the lower band in Fig. \ref{figbarriermod}(c), we have $T^1_{1}+R^1_{1}+T^1_{2}+R^1_{2}+T^1_{3}+R^1_{3}=1.$
\begin{figure*}[tbh]
\centering
\includegraphics[width=6 in]{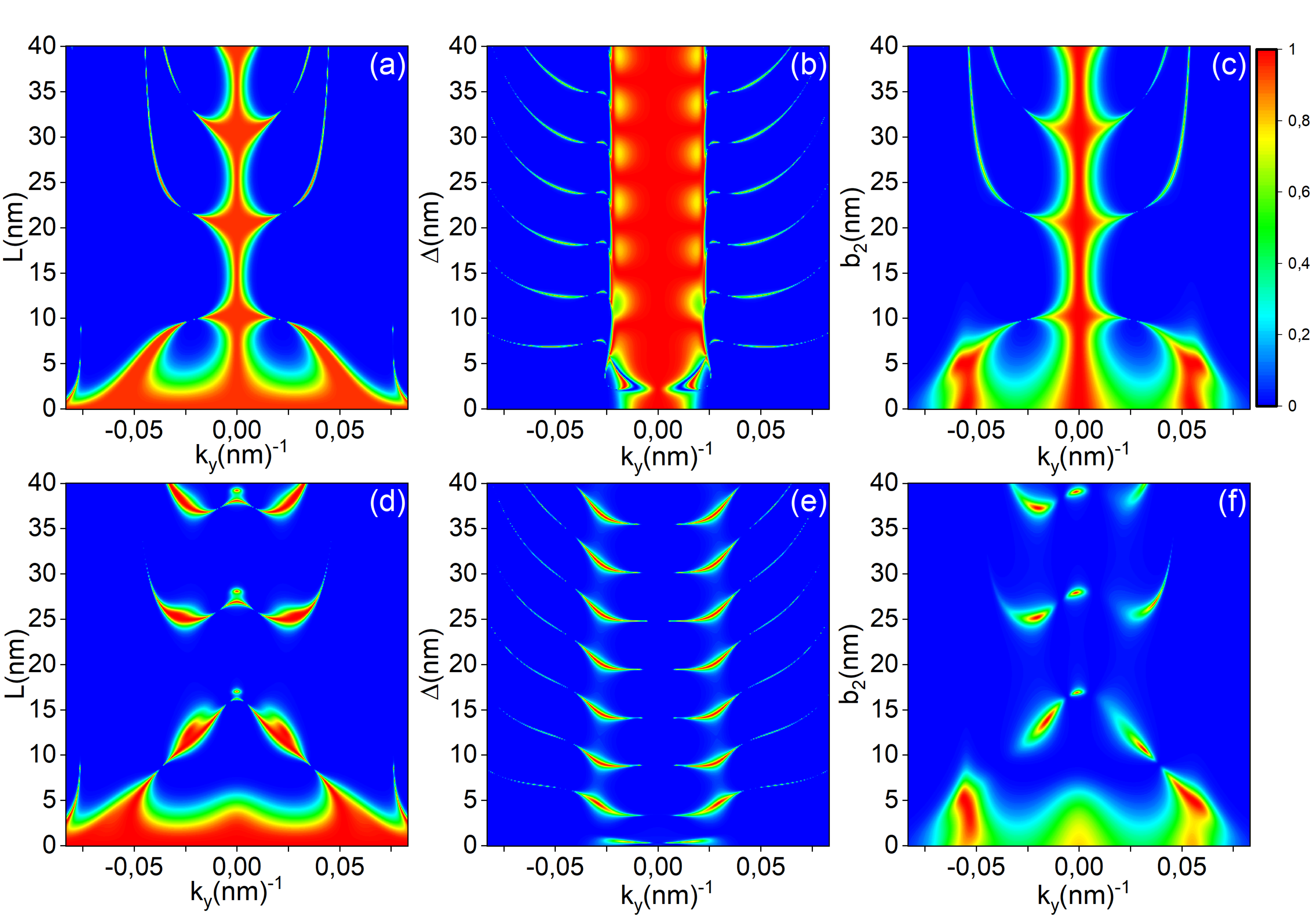}
\caption{(Color online) (Top row) Density plot of the transmission probability $T^1_1$  for $V_2=V_4=0.6\ \gamma_1$, $E=\frac{4}{5}\ V_2$ and for the gap $\delta_2=\delta_4=0$ versus (a): $k_y$ and the width of the two barriers $(b_1=b_2=L)$ and $ \Delta=10$ nm, (b): $k_y$ and $\Delta$ for $b_1=b_2=10$ nm, (c): $ky$ and $b_2$
with $b_1=5$ nm and $\Delta=10$ nm. (Bottom row) The same as top row but for the gap $\delta_2=\delta_4=0.1\ \gamma_1$.}\label{TransLb2Delta}
\end{figure*}

Since we have found transmission probabilities, let see how these will effect
the conductance of our system.  This actually can be obtained
through the Landauer-B\"{u}ttiker formula \cite{Blanter336} by summing
on all channels to end up with
\begin{equation}\label{eq24}
G(E)=G_{0}\frac{L_y}{2
\pi}\int_{-\infty}^{+\infty}dk_{y}\sum^3_{s,n=1}T^{s}_{n}(E,k_y),
\end{equation}
with  the width $L_y$ of our system in the $y$-direction and the conductance unit
$G_0=4e^2/h$, the factor $4$ refers  to the valley and
spin degeneracy in graphene. 

The obtained results will be numerically analyzed to discuss the
basic features of
our system and also make link with other published results. Because of the nature of our system,
we do our task by  distinguishing two different cases
in terms of the band tunneling.

\section{Band tunneling analysis}
\label{tunneling}

We distinguish 
two interesting cases resulted from our energy
spectrum. Indeed,  we will separately treat each case 
 and underline their relevant properties. We will compare our results with previous work\cite{vanduppen195439,Benlakhouy114835,vanduppen205427,ElMouhafid2017}.
\subsection{Tunneling at low energy}

Fig. \ref{TransbasseE} shows a few contour plots of the transmission probability $T^1_1$ corresponds to the propagation from the channel $k^1$ in region I to the channel $k^5=k^1$ in region V at low energy as a function of the 
wave vector $k_y$ of the incident  energy $E$ under suitable conditions of the physical parameters.
We address two different cases: (i) the
height of the two barriers is the same ($V_2=V_4$) (symmetric double barrier structure) and (ii) with different height of the two barriers ($V_2<V_4$) (asymmetric double barrier structure)  with and without interlayer potential on both barriers in both cases. Indeed, the case (i) when $V_2=V_4=0.4\gamma_1$ in Fig. \ref{TransbasseE}(a,b) with $\delta_2=\delta_4=0$ in Fig. \ref{TransbasseE}(a) and $\delta_2=\delta_4=0.2\gamma_1$ in Fig. \ref{TransbasseE}(b) and the the case (ii) when $V_2=0.4\gamma_1$ and $V_4=0.6\gamma_1$ in Fig. \ref{TransbasseE}(b,c) with $\delta_2=\delta_4=0$ in Fig. \ref{TransbasseE}(c) and $\delta_2=\delta_4=0.2\gamma_1$ in Fig. \ref{TransbasseE}(d). It is interesting to note that the Van Duppen et \textit{al.} results \cite{vanduppen195439} can be derived  from our analysis by considering  (i) and requiring $b=c$ in our double barrier structure. However, the effect of the different structure of the two barriers should appear in the transmission and reflection.
\begin{figure*}[tbh]
\begin{center}
\end{center}
\includegraphics[width=18cm]{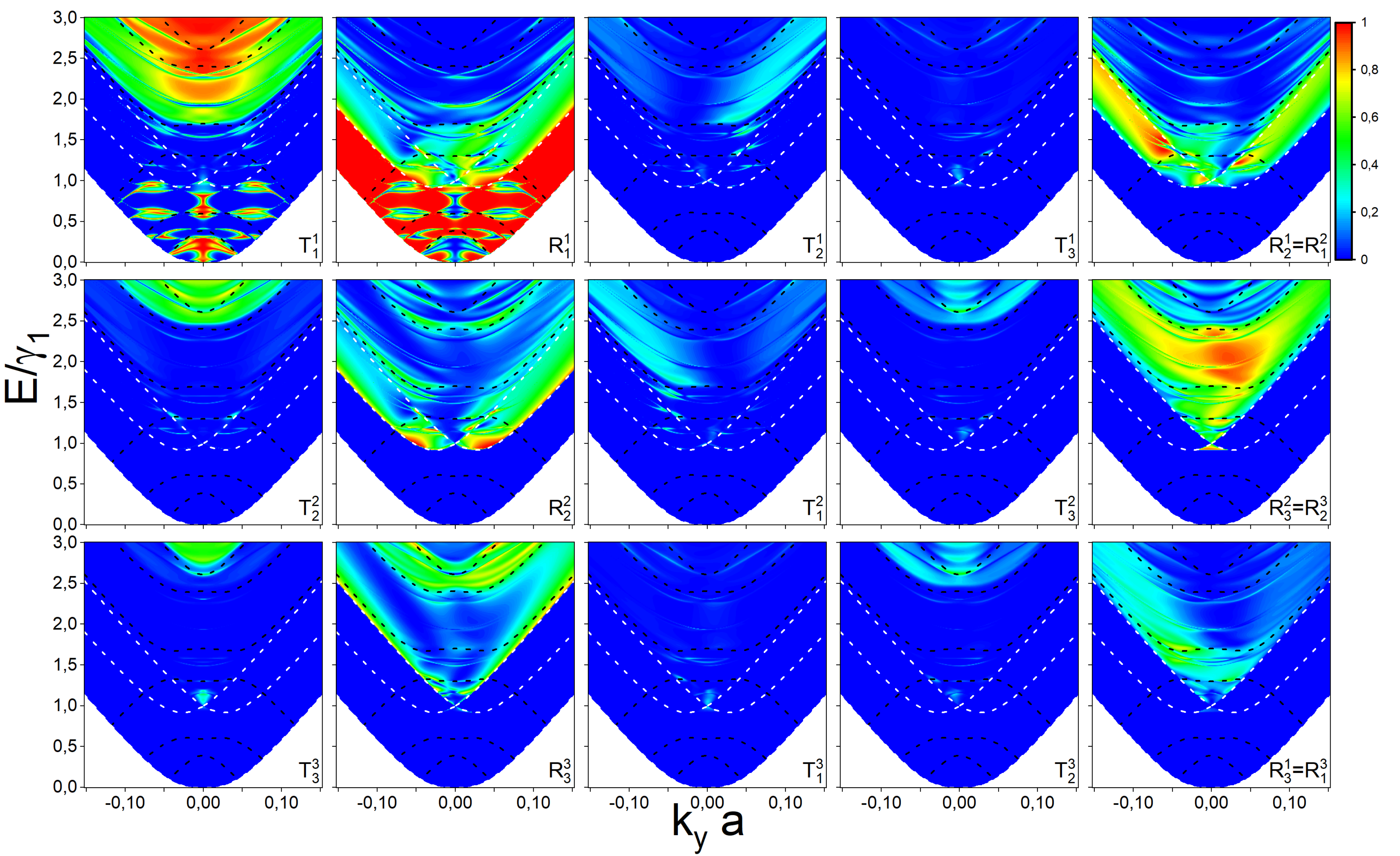}
\caption{(Color online) Density plot of transmission and reflection probabilities
	versus $E$ and $k_y$ with $V_{2}=V_{4}=1.5\ \gamma_{1}$, $\delta_2=\delta_4=0.2\ \gamma_1$ and $b_1=b_2=\Delta=10$ nm. The dashed black lines show the band inside the barrier whereas the white lines represent the band outside
	the barrier.
	.}
\label{TandRSymmetricHE}
\end{figure*}
\begin{figure*}[tbh]
\begin{center}
\end{center}
\includegraphics[width=18cm]{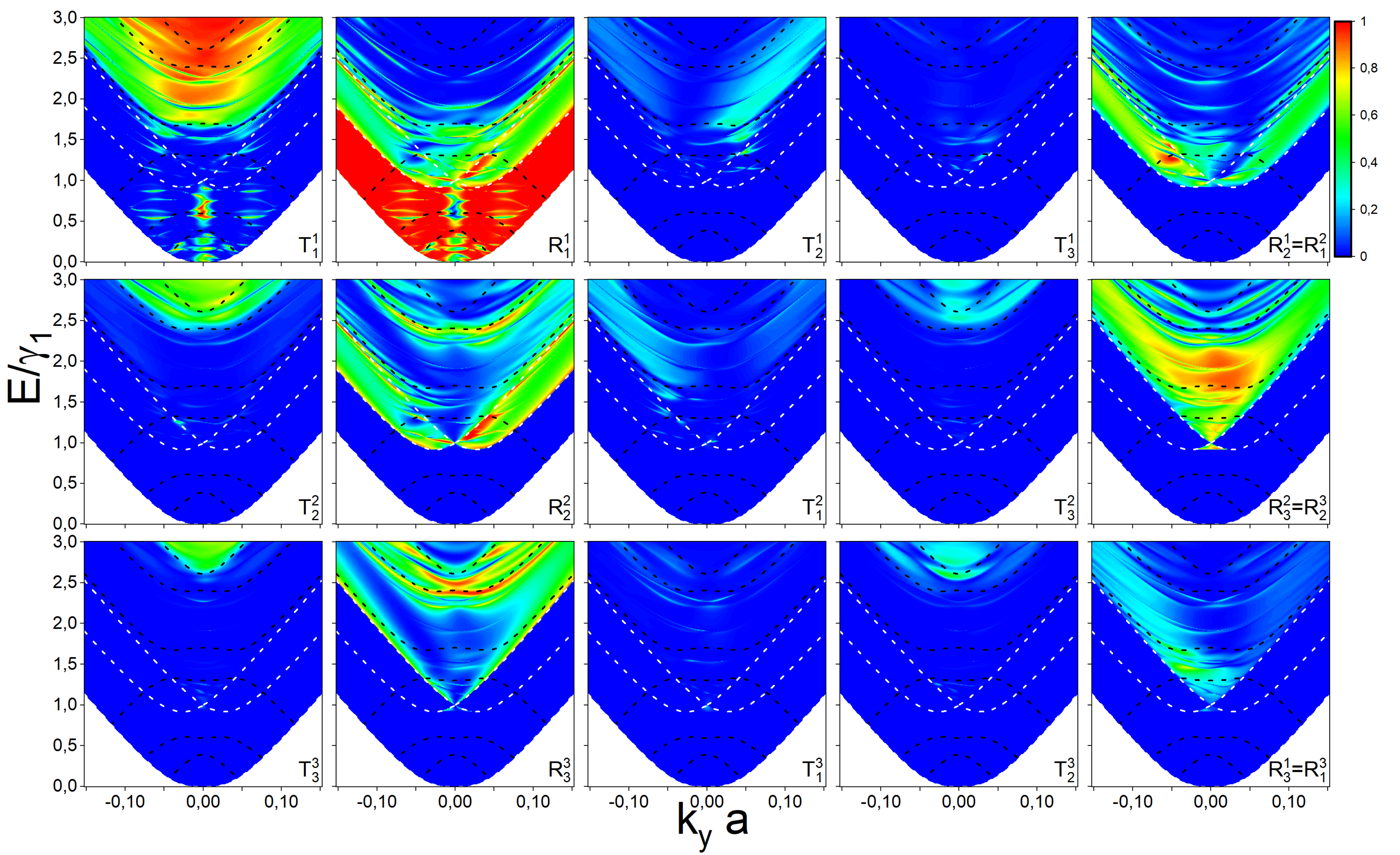}
\caption{(Color online) The same as Fig. \ref{TandRSymmetricHE} but now for $V_{2}=1.3\gamma_{1}$, $V_{4}=1.5\ \gamma_{1}$.}
\label{TandRAsymmetricHE}
\end{figure*}
\begin{figure}[tbh]
\begin{center}
\end{center}
\includegraphics[width=8.74cm]{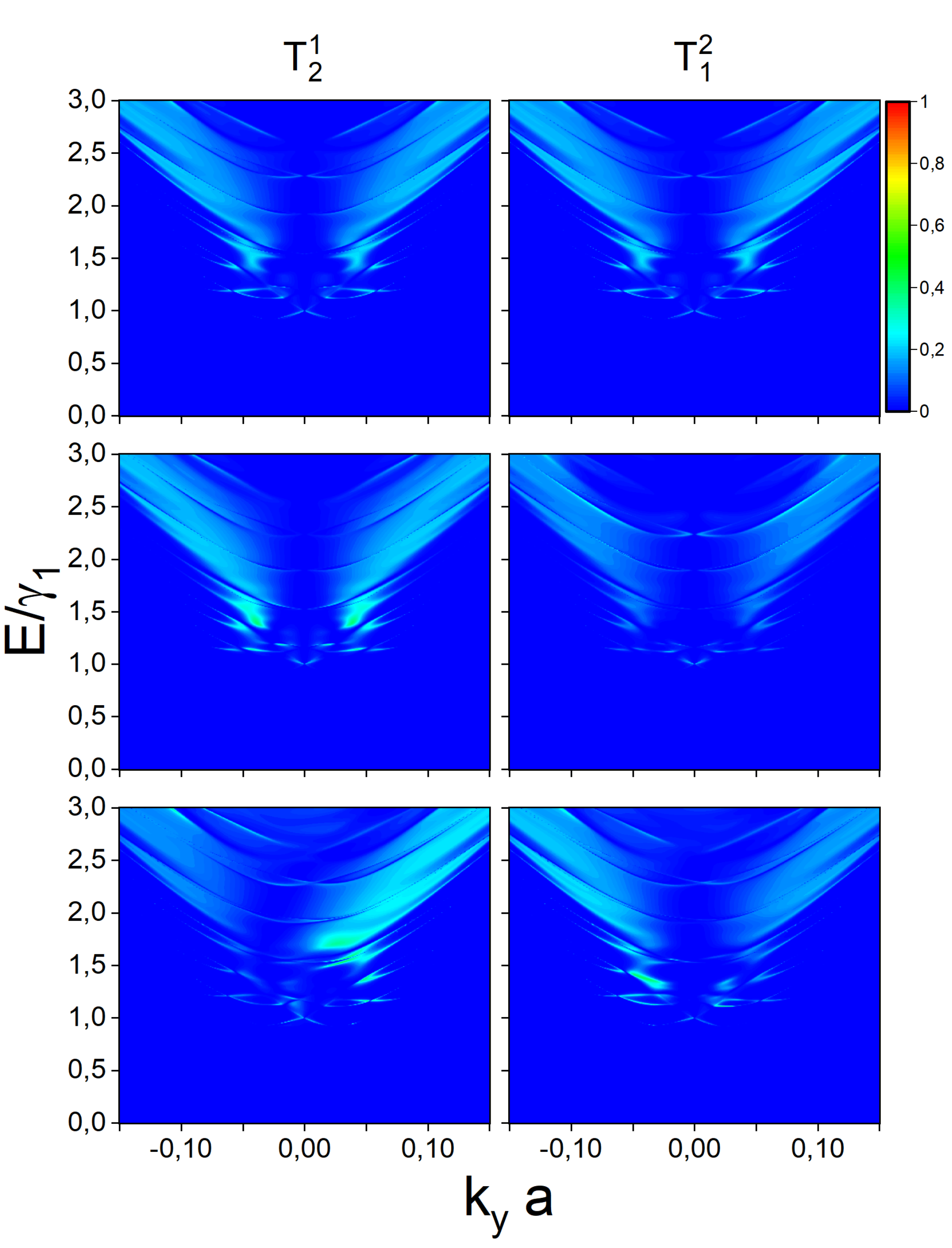}
\caption{(Color online) Density plot of transmission probabilities $T^1_2$ (left panel) and $T^2_1$ (right panel) versus $E$ and $k_y$ with $b_1=b_2=\Delta=10$ nm for   $V_{2}=V_{4}=1.5\ \gamma_{1}$ and $\delta_2=\delta_4=0$ (top row),  $V_{2}=1.3\ \gamma_{1}$, $V_{4}=1.5\ \gamma_{1}$ and $\delta_2=\delta_4=0$ (middle row),  $V_{2}=V_{4}=1.5\ \gamma_{1}$, $\delta_2=0$ and $\delta_4=0.2\gamma_1$ (bottom row).}
\label{Transmissioneky3}
\end{figure}
\begin{figure}[tbh]
\centering
\includegraphics[width=3.2in]{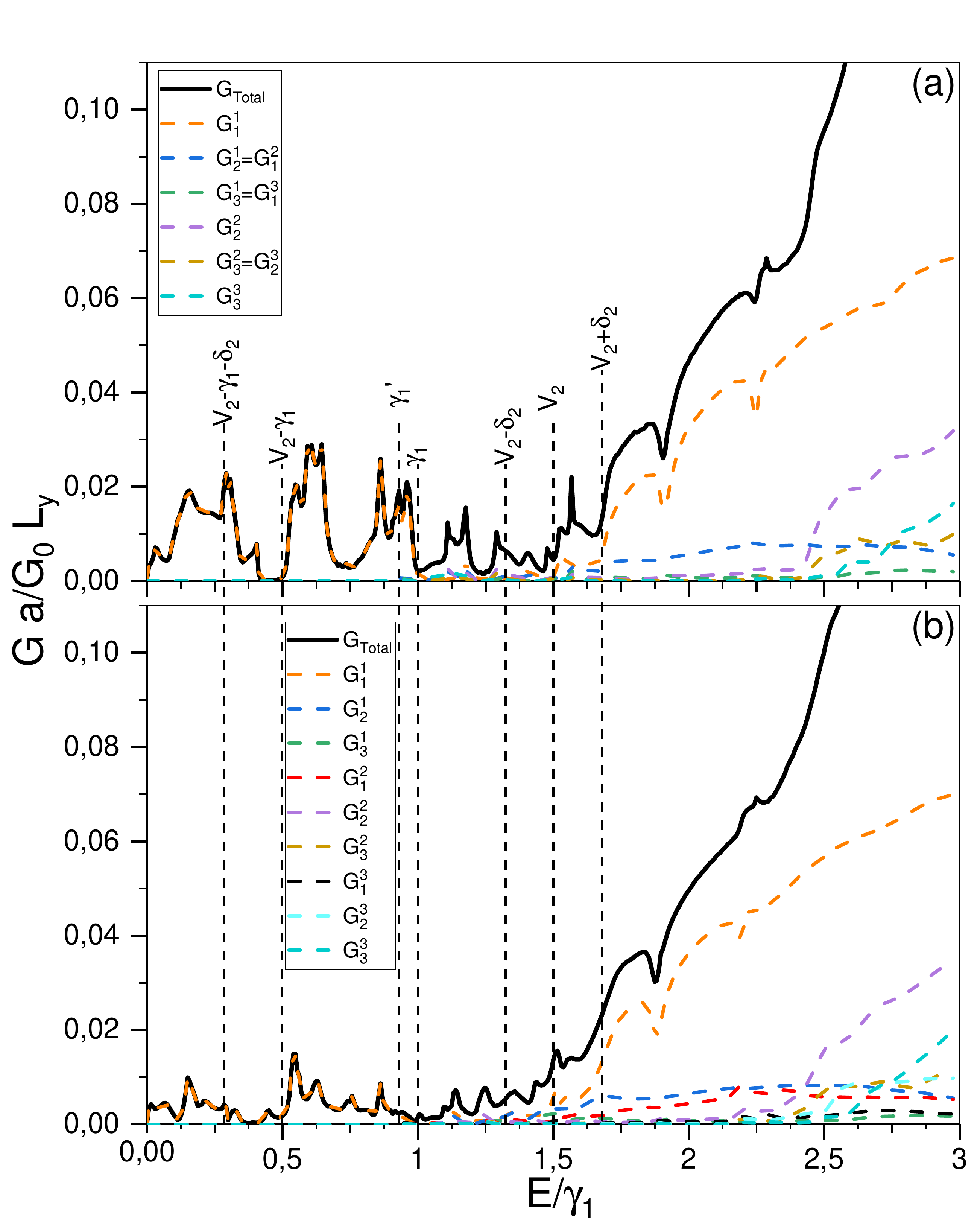}
\caption{(Color online) Conductance for ABC-TLG 
	versus the incident energy $E$ with $b_1=b_2=\Delta=10$ nm and $\delta_2=\delta_4=0.2\gamma_1$. (a): $V_2=V_4=1.5\gamma_1$. (b): $V_{2}=1.3\gamma_1$, $U_{4}=1.5\gamma_1$. The solid lines are for the total conductance and the dashed ones are for the contributions of different propagation channels.}\label{Conductance1}
\end{figure}
\hspace{-.8cm}As shown in Fig. \ref{TransbasseE}(a,c), for a normal incidence ($k_y=0$) and  $\delta_2=\delta_4=0$ the transmission is unit and becomes independent of energy or the barrier structure (symmetric or asymmetric), which is the same from the single barrier case\cite{vanduppen195439}.  This is a manifestation of the Klein tunneling that is resulted from the conservation of pseudospin and occurs for rhombohedrally stacked multilayers with an odd number of layers\cite{vanduppen195439}. For $\delta_2=\delta_4=0.2\ \gamma_1$, we show  $V_2=V_4=0.4\ \gamma_1$ in Fig. \ref{TransbasseE}(b) and  $V_2=0.4\ \gamma_1$, $V_4=0.6\
\gamma_1$ in Fig. \ref{TransbasseE}(d). For single barrier, in  AB-BLG\cite{vanduppen205427,Benlakhouy114835} and ABC-TLG\cite{vanduppen195439}, there are no resonant inside the induced gap in contrary to the double barrier as clearly seen  in Fig. \ref{TransbasseE}(b,d). Without $\delta_2$, $\delta_4$ and for  $k_y\neq 0$ we still have a full transmission (very narrow resonances), even for   $E< V_j$,
which are symmetric in $k_y$. Such resonances get reduced and even disappeared in Fig. \ref{TransbasseE}(b,d) due to the asymmetric structure of double barrier. We notice that the asymmetric structure of  double barrier reduces these resonances resulted from the bound electrons in the well between
the two barriers, which are similar to those obtained
for AB-BLG\cite{ElMouhafid2017}.

Fig. \ref{TransLb2Delta} presents the
density plot of the transmission probability as a function of the
 wave vector $k_y$ and the width of two barriers 
 for
$V_2=V_4=0.6\ \gamma_1$ and $E=\frac{4}{5}V_2$,  $\delta_2=\delta_4=0$ and $\delta_2=\delta_4=0.1\gamma_1$ in top and bottom rows, respectively. For $k_{y}=0$ we have a full transmission regardless of thickness $b_1$ and $b_2$ of the two barriers or distance $\Delta$ between them. In contrary, with increasing $L$, $\Delta$ and $b_2$, the transmission probability dramatically decreases however, some resonances still show up as depicted in Fig. \ref{TransLb2Delta}(a,b,c). The transmission probability in Fig. \ref{TransLb2Delta}(b) is completely different compared to Fig. \ref{TransLb2Delta}(a,c) where the position and number of resonant change. We can see clearly in Fig. \ref{TransLb2Delta}(b) for a gapless ABC-TLG the existence of a  brighter region corresponding  to higher transmission probability for a wide range of $k_y$ between $\pm0.025\ \text{nm}^{-1}$ for any values of barrier well $\Delta$.  This tells us that the crucial parameter in determining the number of resonant peaks together with theirs positions is  the well width $\Delta$ in similar way  to the AB-bilayer graphene\cite{ElMouhafid2017}.

Notice that at non normal incidence, i.e., $ky\neq0$ and for $\delta_2=\delta_4=0$ the transmission still equals unity as a result of the Febry-P\`erot oscillations, independent of width $b_1$ or $b_2$ or well $\Delta$ between the two barriers which is the same from the single barrier case in AB-BLG\cite{vanduppen205427,Benlakhouy114835} and ABC-TLG\cite{vanduppen195439}.

For $\delta_2=\delta_4=0.2\ \gamma_1$, the density plots in the Fig. \ref{TransLb2Delta}(bottom row) and for the chosen symmetric barrier show that Klein tunneling is suppressed at normal incidence for some values of barrier width $b1$ or $b2$ and well $\Delta$, there by illustrating some possible reflection even for zero angle of incidence $\phi=\arcsin{(ky/E)}$. We notice that most of the resonances disappeared and splitted due to the band gap in the energy spectrum generated  from the induced electric field. A full transmission frequently occur for normal and non normal incidence as $L$, $\Delta$ and $b_2$ increases as shown in Fig. \ref{TransLb2Delta} for both gap and gapless ABC-TLG.

\subsection{Tunneling at high  energy}
At high energy where all propagation modes are taken into account, the transmission and reflection probabilities as a function of the incident  energy  $E$ and the wave vector $k_y$, corresponding to the modes schematized in Fig. \ref{figbarriermod}(c), are presented in Fig. \ref{TandRSymmetricHE} and in Fig. \ref{TandRAsymmetricHE} through symmetric $V_2=V_4=1.5\gamma_1$ and asymmetric $V_2=1.3\gamma_1$ and $V_4=1.5\gamma_1$ double barrier structure, respectively, with an interlayer potential difference $\delta_2=\delta_4= 0.2\gamma_1$ and width $b_1=b_2=\Delta=10\ nm$ in both cases. The dashed black lines show the band inside the barrier whereas the white lines represent the band outside
the barrier. These lines separate between different regions in the transmission and reflection probabilities, which can be explained by identifying which modes are propagating inside and outside  our system in Fig. \ref{figbarriermod}(a). In our double barrier potential we have nine channels of transmission rather than six unlike the case of ABC-TLG\cite{vanduppen195439} in one barrier. Specifically, in  double barrier structure the symmetry in the scattered transmission $T^i_j$, with $i\neq j$ and $(i,j)=(1, 2, 3)$, from band $k^i$ to band $k^j$ is broken with respect to the normal incidence $k_y=0$ and $T^i_j(\pm k_y)=T^j_i(\mp k_y)$ for $V_2=V_4$ as shown in Fig. \ref{TandRSymmetricHE} and $T^i_j\neq T^j_i$ for $V_2\neq V_4$ as shown in \ref{TandRAsymmetricHE} as in the case of AB-BLG in \cite{vanduppen205427, Benlakhouy114835} and unlike the case of ABC-TLG\cite{vanduppen195439} in one barrier. This can be also understood by pointing out that for the symmetric (asymmetric) structure the particles scattered from top layer to bottom one once moving from left to right in the  valley $K$ are equivalent (non equivalent) to particles scattered from bottom layer to top one once moving oppositely in the second valley $K'$. However, the element responsible for this asymmetry in transmission is the introduction of an interlayer potential difference $\delta_2$ or $\delta_4$ or both $\delta_2$ and $\delta_4$ in the system as shown in addition to Fig. \ref{TandRSymmetricHE} and Fig. \ref{TandRAsymmetricHE} in Fig. \ref{Transmissioneky3}. For $\delta_2=\delta_4=0$ for symmetric (asymmetric) double barrier structure the symmetry in the scattered transmission $T^i_j$ still valid and $T^i_j= T^j_i$ ($T^i_j\neq T^j_i$) as shown in Fig. \ref{Transmissioneky3} for $T^1_2$ and $T^2_1$. We note that, identically to the transmission probabilities $T^i_j$ analysis, we can clearly see that the behavior of the reflection probabilities $R^i_j$ $((i,j)=1, 2, 3$) is the same in terms of symmetry and asymmetry except that the  $R^i_j(\pm k_y)$ and  $R^j_i(\pm k_y)$ are always equals regardless the existence or not of the interlayer potential difference as in one barrier\cite{vanduppen195439}. The transmissions probabilities $T^i_i$ ($i=1,2,3$) exhibit a very similar behavior with respect to the normal incidence $k_y=0$, which is due to the symmetry of the system unlike the $V1\neq V_4$ the $T^i_i$ are not symmetric and $T^i_i\neq T^j_j$ ($i\neq j$). We observe that The barrier heights act by   reducing the
transmission probabilities. However, theirs effects become more
intense inside the gap, which are due to the fact that the available states
outside the first barrier  are in the same energy zone of the
gap on the second barrier.

At normal incidence, and in contrast to the single barrier case\cite{vanduppen195439} as shown in Fig. \ref{TandRSymmetricHE} and \ref{TandRAsymmetricHE} there are a transmission in $T^1_1$ different than zero inside the gap in the energy spectrum resulted from the available states in the well between the barriers, which are similar to those obtained for AB-BLG\cite{ElMouhafid2017}. In addition, Klein tunneling in $T^1_1$ occur for the range of energy $V_2-\gamma_1<E<\gamma'_1$, $E<V_2-\gamma_1-\delta_2$ and $E>V_2+\delta_2$ for both symmetric and non symmetric barrier where $\gamma_1'=0.918 \gamma_{1}$\cite{vanduppen195439}. Outside the previous energy ranges the transmission $T^1_1$ is nearly zero. For $E<\gamma_1$ the transmission probabilities $T^i_j=0$ $(i\neq j)$ $((i,j)=1, 2, 3$) and the reflection probabilities  $R^1_1\neq0$ and $R^i_j=0$ $(i\neq j)$. It implies that at low energy the propagation from region I to region V is only valid by one channel $k^1$, as in the case of the one barrier in AB-BLG\cite{vanduppen205427, Benlakhouy114835} and ABC-TLG\cite{vanduppen195439}. We observe the suppression in transmission due to cloaking effect at non-normal incidence\cite{vanduppen195439,Rudner156603}, which  
also exists for some states 
as a result of the available  states in the well.
\section{Conductance}
\label{Conductance}
The energy dependence of the conductance of ABC-TLG through symmetric and asymmetric double barrier structure is shown in Figs. \ref{Conductance1}(a) and (b), respectively for different values of the applied gate voltage. The solid lines are  the total conductance and the dashed ones are for the contributions of  different propagation channels. The resonances that are visible in the transmission probabilities in Figs. \ref{TandRSymmetricHE} and \ref{TandRAsymmetricHE}  due to the existence of the bound electron states in the well mentioned in the previous section are responsible for the appearance of the peaks  in the double barrier conductance. These resonances are more important in Fig. \ref{Conductance1}(a) for symmetric structure than in Fig. \ref{Conductance1}(b) for asymmetric structure. Several local maxima and minima are observed which are strongly dependent on the structure of the barrier. It is  clearly seen in Figs. \ref{Conductance1}(a) and (b) that for  $E<\gamma_1'$ once the propagating states become available, the conductance shows an sharply increase to unity in the regime where only one band $T^1_1$ is available for contributing to the conductance. For $E>\gamma_1'$ when more bands become possible, additional transmission channels contribute by
increasing the conductance  to almost perfect one with $G^1_2=G^2_1$, $G^1_3=G^3_1$ and $G^2_3=G^3_2$ in Fig. \ref{Conductance1}(a) for symmetric structure and unlike in the Fig. \ref{Conductance1}(b) when $G^1_2\neq G^2_1$, $G^1_3\neq G^3_1$ and $G^2_3\neq G^3_2$ for asymmetric structure.
\section{Conclusion}

We have investigated the tunneling effect of 
 electrons 
through symmetric and asymmetric double barrier potential in ABC-TLG system. Using  the six band model,  we have derived the solutions of  energy spectrum  in all regions composing our system.  By matching the eigenspinors  at different interfaces, we have determined all possible channels of the  transmission and reflection coefficients. Based on the symmetric and asymmetric double barrier structure, we have studied the transmissions by specifying two energy zones
 $E<\gamma_1'$ (one propagating mode) and $E>\gamma_1'$ (three propagating modes)
 in addition to 
various values of the barrier heights and widths.

Additionally, we have compared our results with previous work \cite{vanduppen195439} (for
$E<\gamma_1$) and showed that without gap at normal incidence ($k_y=0$) for symmetric or asymmetric structure the transmission equals unity independent of energy or width or well between the two barriers or barrier structure as it was the case for a single barrier\cite{vanduppen195439}. At $k_y\neq 0$ we have seen that the transmission shows a sequence of the resonances in the region $E<V_j$, which is a consequence of the  bounded electrons in the well between two barriers. We have showed that to 
control the position and the number of
these resonances, in both cases $E<\gamma_1'$ or $E>\gamma_1'$, it is interesting to use 
the well width between the tow barriers rather than  the thickness of the
barriers as obtained in AB-BLG\cite{ElMouhafid2017}.

It was showed that the asymmetric structure of the double potential barrier reduces the transmission probabilities and removes the sharp resonant peaks as well. Furthermore, the transmission symmetry is prized  and the element responsible for this asymmetry is the presence of an interlayer potential difference in region II or in region IV or both II and IV. However, we have observed that the resulting conductance for the double barrier turned into distinctive from that of the single barrier. This  distinction  manifests itself via the presence of many extra resonances  that are related to  the bound electron states in the well.

\section*{Acknowledgment}
The generous support provided by the Saudi Center for Theoretical Physics (SCTP) is highly appreciated by all authors


\end{document}